\documentstyle[prd,preprint,tighten,aps,eqsecnum,amssymb,newlfont,epsfig,floats]{revtex}
\begin{document}
\preprint{
\begin{tabular}{r}
UWThPh-2000-11\\
October 2000
\end{tabular}}

\vspace{2cm}

\title{Schwinger terms in two-dimensional gravitation and K\"all\'en's method\thanks{This work was partly
supported by Austria-Czech Republic Scientific collaboration, project KONTACT 1999-8.}}

\vspace{2cm}

\author{R.A. Bertlmann and E. Kohlprath\thanks{Supported by a Wissenschaftsstipendium der
Magistratsabteilung 18 der Stadt Wien.}}
\address{Institut f\"ur Theoretische Physik,
Universit\"at Wien\\
Boltzmanngasse 5,
A-1090 Vienna, Austria}

\maketitle

\begin{abstract}
We evaluate the gravitational Schwinger terms for the specific 2-dimensional model of Weyl
fermions in a gravitational background field using a technique introduced by K\"all\'en and
find a relation which connects the Schwinger terms with the linearized gravitational anomalies.
\end{abstract}

\section{Introduction}

The appearance of Schwinger terms \cite{Schwinger51} -- \cite{GotoImamura} in the equal time commutators (ETC) of currents is an indication that the corresponding quantum field theory may have anomalies (for an introduction see Refs. \cite{Jackiw1} -- \cite{Bertlmann}). These Schwinger terms (ST) show up as extensions in the canonical algebra of the ETC of the Gauss law operators (see e.g. Refs. \cite{Pawlowski98} -- \cite{Sykora}), and they have a cohomological \cite{Faddeev}, \cite{Mickelsson} and geometrical \cite{NiemiSemenoff} -- \cite{Ekstrand} interpretation. Within QFT the perturbative calculations using the Bjorken-Johnson-Low limit \cite{Bjorken}, \cite{JohnsonLow} work quite successfully, however, not all of the familiar point splitting methods lead to the correct result \cite{BertlmannSykora} -- \cite{Jo}. Therefore it is of interest to shed some light onto the anomalous phenomena from a different point of view. We want to present in our paper a method which is computational quite easy and has interesting features. It has been originally introduced by K\"all\'en \cite{Kaellen} and is related to the dispersion relation approach.  S\'ykora \cite{Sykora} has applied the method to compute the ST for currents in Yang-Mills theories. Our aim is to generalize this procedure to the case of gravitation, where the current is replaced by the energy-momentum tensor, and we perform the calculations in two dimensions. So we have to evaluate the vacuum expectation value of ETC of the energy-momentum tensors.

\section{Schwinger terms}

We start with the Lagrangian describing a Weyl fermion in a gravitational background field in two dimensions

\begin{equation} \label{Lagrangian}
\mathcal{L} = i e E^{a \mu} \bar \psi \gamma_{a} \frac{1}{2} \stackrel{\leftrightarrow}{D}_{\mu}P_{\pm}  \psi, \qquad P_{\pm}=\frac{1\pm\gamma_{5}}{2},
\end{equation}

where $E^{a \mu}$ is the  inverse zweibein, $e = |det\, e^{a}_{\,\,\mu}|$ is the determinant of the zweibein and $D_{\mu} = \partial_{\mu} + \omega_{\mu}$ is the covariant derivative with the spin connection $\omega_{\mu}$. Note that as gravity is only used as an external field we need not specify which theory of gravity we actually mean. The computation will be valid for any two dimensional model where gravity is described by a zweibein. (Remember in two dimensions Einstein's gravity is only topological.) 

From Eq. (\ref{Lagrangian}) we obtain the following classical energy-momentum tensor

\begin{equation} \label{emt}
T_{\mu\nu}=\ :\frac{i}{4}\bar\psi\left(\gamma_{\mu} \stackrel{\leftrightarrow}{\partial}_{\nu}+\gamma_{\nu} \stackrel{\leftrightarrow}{\partial}_{\mu}\right) P_{\pm}\psi:\ =\frac{1}{2}\left( T_{\mu\nu}^{V}\pm T_{\mu\nu}^{A}\right),
\end{equation}

where $\gamma_{\mu}(x)=e^{a}_{\ \mu}(x)\gamma_{a}$. We use the following conventions in two dimensions: $g_{00}=-g_{11}=1$, $\varepsilon^{01}=1$, $\gamma^{0}=\sigma^{2}$, $\gamma^{1}=i\sigma^{1}$ and $\gamma_{5}=\gamma^{0}\gamma^{1}=\sigma^{3}$, where $\sigma^{i}$ are the Pauli matrices.\\

Using the relation

\begin{equation} \label{elimination of gamma5}
\gamma_{\mu}\gamma_{5} = -\varepsilon_{\mu \nu} \gamma^{\nu},
\end{equation}

and the equations of motions we can express the pseudotensor part of the energy-momentum tensor by the pure tensor part (recall that the tensor is symmetric)

\begin{equation}
T_{\mu\nu}^{A}= -\varepsilon_{\mu}^{\ \lambda}T_{\lambda\nu}^{V}\,.
\end{equation}

In quantum field theory the energy-momentum tensors form an algebra which is generally not closed but has central extensions, socalled Schwinger terms. In two dimensions we have

\begin{eqnarray}
\lbrack T_{00}(x), T_{00}(0) \rbrack_{ET} &=& i \left(T_{01}(x) + T_{01}(0) \right) \partial_{1}\delta(x^{1}) + S_{0000}\\ 
\lbrack T_{01}(x), T_{01}(0) \rbrack_{ET} &=& i \left(T_{01}(x) + T_{01}(0) \right) \partial_{1}\delta(x^{1}) + S_{0101}\\ 
\lbrack T_{00}(x), T_{01}(0) \rbrack_{ET} &=& i \left(T_{00}(x) + T_{00}(0) \right) \partial_{1}\delta(x^{1}) + S_{0001} \, . 
\end{eqnarray}

The Schwinger terms $S_{0000}, S_{0101}, S_{0001}$ are c-number terms and can therefore be determined by
considering the vacuum expectation value of the ETC.

\section{K\"all\'en's method}

To evaluate the vacuum expectation value of the ETC we work with  a technique that has been introduced by K\"all\'en
\cite{Kaellen} and is closely related to the dispersive approach. This technique
has been applied already by S\'ykora \cite{Sykora} to compute the ST for currents in Yang-Mills
theories. Our aim is to generalize this procedure to the case of gravitation, where the current
is replaced by the energy-momentum tensor.\\

In two dimensions we have the identity

\begin{equation}
\varepsilon_{\mu}^{\ \lambda}\varepsilon_{\rho}^{\ \tau}=-g_{\mu\rho}g^{\lambda\tau}+g^{\lambda}_{\ \rho}g^{\tau}_{\ \mu}\,,
\end{equation}

so that we find

\begin{eqnarray} \label{symbolic VEV of commutator}
&&\langle 0\vert [T_{\mu\nu}(x),T_{\rho\sigma}(0)]\vert 0\rangle = \frac{1}{4} \Biggl\{ \langle 0\vert [T^{V}_{\mu\nu}(x),T^{V}_{\rho\sigma}(0)]\vert 0\rangle {}\nonumber\\{}&& \qquad+ \langle 0\vert [T^{V}_{\rho\nu}(x),T^{V}_{\mu\sigma}(0)]\vert 0\rangle - g_{\mu\rho} \langle 0\vert [T^{V\lambda}_{\nu}(x),T^{V}_{\lambda\sigma}(0)]\vert 0\rangle {}\nonumber\\{}&& \qquad \mp \varepsilon_{\mu}^{\ \lambda}\langle 0\vert [T^{V}_{\lambda\nu}(x),T^{V}_{\rho\sigma}(0)]\vert 0\rangle \mp \varepsilon_{\rho}^{\ \lambda} \langle 0\vert [T^{V}_{\mu\nu}(x),T^{V}_{\lambda\sigma}(0)]\vert 0\rangle \Biggr\}.
\end{eqnarray}

So it is enough to consider the commutator of the pure tensor parts.\\

Let us define the following pure tensor contribution

\begin{equation}
F_{\mu\nu\rho\sigma}(x):= \langle 0\vert T^{V}_{\mu\nu}(x) T^{V}_{\rho\sigma}(0) \vert 0\rangle.
\end{equation}

By inserting a complete set of states $\vert n\rangle$ with positive energy and momentum $p_{n}$ and using the translation invariance we obtain

\begin{eqnarray}
F_{\mu\nu\rho\sigma}(x)&=& \sum_{n}\langle 0\vert T_{\mu\nu}^{V}(x) \vert n\rangle\langle n\vert T_{\rho\sigma}^{V}(0)\vert 0\rangle {}\nonumber\\{}
&=& \sum_{n}\langle 0\vert T_{\mu\nu}^{V}(0) \vert n\rangle\langle n\vert T_{\rho\sigma}^{V}(0)\vert 0\rangle e^{-ip_{n}x}.
\end{eqnarray}

We may write

\begin{equation}
F_{\mu\nu\rho\sigma}(x) = \int\!d^{2}p\ e^{-ipx}G_{\mu\nu\rho\sigma}(p)\theta(p^{0}),
\end{equation}

where

\begin{equation} \label{symbolic Gmunurhosigma}
G_{\mu\nu\rho\sigma}(p)=\sum_{n}\delta(p_{n}-p)\langle 0\vert T_{\mu\nu}^{V}(0) \vert n\rangle\langle n\vert T_{\rho\sigma}^{V}(0)\vert 0\rangle.
\end{equation}

From Lorentz covariance and symmetry we get the following decompositon into formfactors 

\begin{eqnarray}
G_{\mu \nu \rho \sigma}(p) &=& p_{\mu}p_{\nu}p_{\rho}p_{\sigma} G_{1}(p^{2}) + (p_{\mu}p_{\nu} g_{\rho \sigma} + p_{\rho}p_{\sigma} g_{\mu \nu}) G_{2}(p^{2}) {} \nonumber \\ {} & &+ (p_{\mu}p_{\rho} g_{\nu \sigma} + p_{\mu}p_{\sigma} g_{\nu \rho} + p_{\nu}p_{\rho} g_{\mu \sigma} + p_{\nu}p_{\sigma} g_{\mu \rho}) G_{3}(p^{2}) {} \nonumber \\ {} & &+ g_{\mu \nu}g_{\rho \sigma} G_{4}(p^{2}) + (g_{\mu \rho}g_{\nu \sigma} + g_{\mu \sigma}g_{\nu \rho}) G_{5}(p^{2}) \, .
\end{eqnarray}

Making use of $\partial^{\mu} T^{V}_{\mu\nu}(x)=0$ provides the Ward identity $p^{\mu}G_{\mu\nu\rho\sigma}(p)=0$ that can be expressed by the formfactors in the following way

\begin{eqnarray}
p^{2}G_{1}+G_{2}+2G_{3}&=&0 \label{VWI1 for G}\\
p^{2}G_{2}+G_{4}&=&0 \label{VWI2 for G}\\
p^{2}G_{3}+G_{5}&=&0 \label{VWI3 for G}.
\end{eqnarray}

Now let us explicitly evaluate $G_{\mu\nu\rho\sigma}(p)$. As we are considering the energy-momentum tensor as a free (noninteracting) tensor -- analogously to the case of free currents -- we only need to sum up states that consist of one fermion-antifermion pair in Eq. (\ref{symbolic Gmunurhosigma}). We get

\begin{equation}
G_{\mu\nu\rho\sigma}(p)=\int\!dp_{1}\int\!dp_{2}\sum_{s_{1}}\sum_{s_{2}}\ \delta(p-p_{1}-p_{2})\langle 0\vert T_{\mu\nu}^{V}(0) \vert p_{1},s_{1};p_{2},s_{2}\rangle\langle p_{1},s_{1};p_{2},s_{2}\vert T_{\rho\sigma}^{V}(0)\vert 0\rangle.
\end{equation}

Let us assume that the fermions are described by a canonically quantized field with mass $m$. We then find with $(/\hspace{-6.5pt}p -m)u(p)=0$ and $(/\hspace{-6.5pt}p +m)v(p)=0$

\begin{eqnarray}
G_{\mu\nu\rho\sigma}(p) &=&-\frac{1}{64 \pi^{2}} \int\!dp_{1}\int\!dp_{2}\sum_{s_{1}}\sum_{s_{2}}\ \delta(p-p_{1}-p_{2})\frac{m^{2}}{E_{p_{1}}E_{p_{2}}}{}\nonumber\\{}&&\times  (p_{1}-p_{2})_{\nu}(p_{1}-p_{2})_{\sigma}\bar v^{(s_{2})}(p_{2})\gamma_{\mu}u^{(s_{1})}(p_{1})\bar u^{(s_{1})}(p_{1})\gamma_{\rho}v^{(s_{2})}(p_{2})  {}\nonumber\\{}&&+ (\mu \leftrightarrow \nu) + (\rho \leftrightarrow \sigma) + {\mu \leftrightarrow \nu \choose \rho \leftrightarrow \sigma}.
\end{eqnarray}

Without the interchanges we call this $G^{ni}_{\mu\nu\rho\sigma}(p)$.\\

We then use the completeness relations for the spinors and introduce the $0$-component of the 2-momenta as integration variable to obtain

\begin{eqnarray}
G^{ni}_{\mu\nu\rho\sigma}&=&-\frac{1}{64 \pi^{2}}\int\!d^{2}p_{1}d^{2}p_{2}\ \delta(p-p_{1}-p_{2})\delta(p_{1}^{2}-m^{2})\delta(p_{2}^{2}-m^{2})\theta(p_{1}^{0})\theta(p_{2}^{0}) {}\nonumber\\{}&&\times (p_{1}-p_{2})_{\nu}(p_{1}-p_{2})_{\sigma}tr\Bigl[ \gamma_{\mu}(/\hspace{-6.5pt}p_{1}+m)\gamma_{\rho}(/\hspace{-6.5pt}p_{2}-m) \Bigr].
\end{eqnarray}

Integrating next over the first $\delta$-function and evaluating the trace gives

\begin{eqnarray}
G^{ni}_{\mu\nu\rho\sigma}&=&-\frac{1}{32 \pi^{2}}\int\!d^{2}p_{1}\ \delta(p_{1}^{2}-m^{2})\delta((p-p_{1})^{2}-m^{2})\theta(p_{1}^{0})\theta(p^{0}-p_{1}^{0}) {}\nonumber\\{}&&\times (2p_{1}-p)_{\nu}(2p_{1}-p)_{\sigma}\Bigl[ p_{1\mu}(p-p_{1})_{\rho}+p_{1\rho}(p-p_{1})_{\mu}{}\nonumber\\{}&& -g_{\mu\rho}p_{1}^{\lambda}(p-p_{1})_{\lambda}-g_{\mu\rho}m^{2} \Bigr].
\end{eqnarray}

If we compare this with the following amplitude

\begin{equation} \label{amplitude expressed by T-product}
T^{pv}_{\mu \nu \rho \sigma}(p) = i \int d^{2}x e^{i p x} \langle 0 \vert T [T^{V}_{\mu\nu}(x) T^{V}_{\rho\sigma}(0)] \vert 0 \rangle,
\end{equation}

where ``pv'' stands for ``pure vector'' (representing the pure tensor contribution), we find

\begin{equation} \label{G and Tpv}
G_{\mu\nu\rho\sigma}(p)=\frac{1}{2\pi^{2}} Im T^{pv}_{\mu\nu\rho\sigma}(p) \, .
\end{equation}

So $G_{\mu\nu\rho\sigma}(p)$ is proportional to the imaginary part of the amplitude. From $G_{\mu\nu\rho\sigma}(p)$ we will determine the Schwinger terms whereas from $Im T^{pv}_{\mu\nu\rho\sigma}(p)$ we find $T^{pv}_{\mu \nu \rho \sigma}(p)$ due to dispersion relations (see Ref. \cite{BertlmannKohlprath}).\newpage

Using (\ref{emt}), (\ref{elimination of gamma5}) one then gets the full amplitude

\begin{equation} 
T_{\mu \nu \rho \sigma}(p) = i \int d^{2}x e^{i p x} \langle 0 \vert T [T_{\mu\nu}(x) T_{\rho\sigma}(0)] \vert 0 \rangle.
\end{equation}

From the anomalous Ward identities $p^{\mu}T_{\mu \nu \rho \sigma}(p)$ and $g^{\mu\nu}T_{\mu \nu \rho \sigma}(p)$ then follow the linearized Einstein $\langle\partial^{\mu}T_{\mu\nu}\rangle$ and Weyl $\langle T^{\mu}_{\ \mu}\rangle$ anomalies respectively. (See Ref. \cite{BertlmannKohlprath} for details.) So in this approach  relation (\ref{G and Tpv}) links the Schwinger terms to the gravitational anomalies.\\

The reader may note that the two-point function $T_{\mu \nu \rho \sigma}(p)$ is dependent from the gravitational background field as the energy-momentum tensors in (\ref{emt}) depend on the zweibein. If we now introduce a linearized graviton and do perturbation theory by coupling it to the energy-momentum tensor, $T_{\mu \nu \rho \sigma}(p)$ will give us the leading contribution to the gravitational anomalies. Independent of this linearization $G_{\mu\nu\rho\sigma}(p)$ determines the full Schwinger terms in a gravitational background.\\

Using the expression

\begin{eqnarray}
J_{0} &=& \int \!d^{2}k \ \delta(k^{2}-m^{2}) \delta\Bigl((p+k)^{2}-m^{2}\Bigr) \theta(-k^{0}) \theta(k^{0}+p^{0}) {}\nonumber\\{}&=&  \frac{1}{p^{2}}\left(1-\frac{4m^{2}}{p^{2}}\right)^{\hspace{-3pt}-\frac{1}{2}}\theta(p^{2}-4m^{2})
\end{eqnarray}

we can compute the formfactors (see Ref.\cite{BertlmannKohlprath}, \cite{Kohlprath})

\begin{eqnarray}
G_{1}(p^{2}) &=& -\frac{1}{4\pi^{2}} J_{0}\frac{m^{2}}{p^{2}}\left(1 -4\frac{m^{2}}{p^{2}}\right) \label{explicit G1}\\
G_{2}(p^{2}) &=& -\frac{1}{48\pi^{2}} J_{0}p^{2}\left(1-8\frac{m^{2}}{p^{2}}+16\frac{m^{4}}{p^{4}}\right) \\
G_{3}(p^{2}) &=& \frac{1}{96\pi^{2}} J_{0}p^{2}\left(1+4\frac{m^{2}}{p^{2}}-32\frac{m^{4}}{p^{4}}\right) \\
G_{4}(p^{2}) &=& \frac{1}{48\pi^{2}} J_{0}p^{4}\left(1-8\frac{m^{2}}{p^{2}}+16\frac{m^{4}}{p^{4}}\right) \\
G_{5}(p^{2}) &=& -\frac{1}{96\pi^{2}} J_{0}p^{4}\left(1+4\frac{m^{2}}{p^{2}}-32\frac{m^{4}}{p^{4}}\right). \label{explicit G5}
\end{eqnarray}

Now we consider the commutator

\begin{equation} \label{[T,T] in terms of G}
\langle 0\vert [T^{V}_{\mu\nu}(x), T^{V}_{\rho\sigma}(0)] \vert 0\rangle = F_{\mu\nu\rho\sigma}(x)-F_{\rho\sigma\mu\nu}(-x) = \int\!d^{2}p\ e^{-ipx} \varepsilon(p^{0}) G_{\mu\nu\rho\sigma}(p) \, .
\end{equation}

If we remove the mass, $m \rightarrow 0$, that acted as an infrared cutoff, we get

\begin{equation}
G_{1}(p^{2}) = \lim_{m\rightarrow 0}-\frac{1}{4\pi^{2}}\frac{m^2}{p^4}\left(1-\frac{4m^{2}}{p^{2}}\right)^{\hspace{-3pt}\frac{1}{2}}\theta\left(p^{2}-4m^{2}\right) = -\frac{1}{24\pi^{2}}\delta(p^{2}) \, .
\end{equation}

\newpage

From Eq. (\ref{[T,T] in terms of G}) we explicitly find

\begin{eqnarray}
\langle 0\vert [T^{V}_{00}(x), T^{V}_{00}(0)] \vert 0\rangle_{ET} &=& \lim_{x_{0}\rightarrow 0} -\frac{1}{24\pi^{2}}\int\!d^{2}p\ e^{-ipx}p_{1}^{4}\ \varepsilon(p^{0}) \delta(p^{2}) = 0 \\
\langle 0\vert [T^{V}_{11}(x), T^{V}_{11}(0)] \vert 0\rangle_{ET} &=& \lim_{x_{0}\rightarrow 0} -\frac{1}{24\pi^{2}}\int\!d^{2}p\ e^{-ipx}p_{0}^{4}\ \varepsilon(p^{0}) \delta(p^{2}) = 0 \\
\langle 0\vert [T^{V}_{00}(x), T^{V}_{11}(0)] \vert 0\rangle_{ET} &=& \langle 0\vert [T^{V}_{01}(x), T^{V}_{01}(0)] \vert 0\rangle_{ET} {}\nonumber\\{}&=&  \lim_{x_{0}\rightarrow 0} -\frac{1}{24\pi^{2}}\int\!d^{2}p\ e^{-ipx}p_{0}^{2}p_{1}^{2}\ \varepsilon(p^{0}) \delta(p^{2}) = 0 \\
\langle 0\vert [T^{V}_{00}(x), T^{V}_{01}(0)] \vert 0\rangle_{ET} &=& \lim_{x_{0}\rightarrow 0} -\frac{1}{24\pi^{2}}\int\!d^{2}p\ e^{-ipx}p_{0}p_{1}^{3}\ \varepsilon(p^{0}) \delta(p^{2}) \\
\langle 0\vert [T^{V}_{01}(x), T^{V}_{11}(0)] \vert 0\rangle_{ET} &=& \lim_{x_{0}\rightarrow 0} -\frac{1}{24\pi^{2}}\int\!d^{2}p\ e^{-ipx}p_{0}^{3}p_{1}\ \varepsilon(p^{0}) \delta(p^{2}) \, .\label{[T01,T11] no interchange} 
\end{eqnarray}

The first three expressions vanish because $\varepsilon(p^{0})$ is antisymmetric. To evaluate the next two we use the Pauli-Jordan function 

\begin{equation}
\triangle(x)=\frac{1}{2\pi} \int\!d^{2}p\ e^{-ipx} \varepsilon(p^{0})\delta(p^{2})
\end{equation}

with the properties

\begin{eqnarray}
\partial^{\mu}\partial_{\mu} \triangle(x)&=&0\\
\left. \triangle(x)\right|_{x^{0}=0}&=&0\\
\left. \partial_{0}\triangle(x)\right|_{x^{0}=0}&=&-i\delta(x^{1})
\end{eqnarray}

and we find

\begin{equation}
\lim_{x_{0}\rightarrow 0}-\frac{1}{24\pi^{2}}\int\!d^{2}p\ e^{-ipx}\varepsilon(p^{0}) p_{0}p_{1}p^{2}\delta(p^{2}) = \lim_{x_{0}\rightarrow 0}-\frac{1}{12\pi}\partial_{0}\partial_{1}(\partial_{0}^{2}-\partial_{1}^{2})\triangle(x)=0 \, .
\end{equation}

So we conclude

\begin{eqnarray}
\langle 0\vert [T^{V}_{00}(x), T^{V}_{01}(0)] \vert 0\rangle_{ET} &=& \langle 0\vert [T^{V}_{01}(x), T^{V}_{11}(0)] \vert 0\rangle_{ET} {}\nonumber\\{}&=& \lim_{x_{0}\rightarrow 0}-\frac{1}{12\pi}\partial_{1}^{3}\partial_{0}\triangle(x) = \frac{i}{12\pi}\partial_{1}^{3}\delta(x^{1}) \, .
\end{eqnarray}

With Eq. (\ref{symbolic VEV of commutator}) we finally obtain the Schwinger terms in the ETC of the energy-momentum tensors

\begin{eqnarray}
\langle 0\vert [T_{00}(x), T_{00}(0)] \vert 0\rangle_{ET} &=& \langle 0\vert [T_{11}(x), T_{11}(0)] \vert 0\rangle_{ET} =\mp\frac{i}{24\pi}(\partial_{1})^{3}\delta(x^{1}) \label{ST1} \\ 
\langle 0\vert [T_{00}(x), T_{11}(0)] \vert 0\rangle_{ET} &=& \langle 0\vert [T_{01}(x), T_{01}(0)] \vert 0\rangle_{ET} =\mp\frac{i}{24\pi}(\partial_{1})^{3}\delta(x^{1}) \label{ST2} \\
\langle 0\vert [T_{00}(x), T_{01}(0)] \vert 0\rangle_{ET} &=& \langle 0\vert [T_{01}(x), T_{11}(0)] \vert 0\rangle_{ET} =\frac{i}{24\pi}(\partial_{1})^{3}\delta(x^{1}) \, . \label{ST3}
\end{eqnarray}

\section{Conclusion}

We have calculated the ETC of the energy-momentum tensors by using a technique introduced by K\"all\'en \cite{Kaellen}. Our result, Eqs. (\ref{ST1})--(\ref{ST3}), agrees with the one of Tomiya \cite{Tomiya} who works with a different method, essentially equivalent to the Bjorken-Johnson-Low limit, and in addition uses a cohomological approach. It also coincides with the result of Ebner, Heid and Lopes-Cardoso \cite{EbnerHeidLopes} who derive the ST directly from the gravitational anomaly. In our approach Eq. (\ref{G and Tpv}) is the basic relation. It shows clearly how the Schwinger terms, given by the quantity $G_{\mu\nu\rho\sigma}$, are connected with the gravitational anomalies, as the imaginary part $Im\,T^{pv}_{\mu\nu\rho\sigma}$ determines the gravitational anomalies by means of dispersion relations \cite{BertlmannKohlprath}. We observe that it is just the formfactor $G_{1}(p^{2})$ which contributes to the ST due to its peculiar feature that it approaches a $\delta$-function singularity at zero momentum squared when $m\to 0$. This is the characteristic infrared feature of a dispersion relation approach. It should be noted, however, that this infrared singularity is of different nature as compared to the familiar infrared divergencies in flat space-time. The former arises from massless fermions whereas the later result from massless vector bosons.\\

The main physical implication of the Schwinger terms in the algebra of energy-momentum tensors is the following: If one chooses a specific two dimensional model for gravity, then some of the components of the energy-momentum tensor will be part of the constraints in the Hamiltonian formulation. (See e.g. Ref. \cite{NelsonTeitelboim}). The Schwinger terms in the energy-momentum tensors will then lead to nontrivial Schwinger terms in the constraints giving an obstruction to a consistent quantization of the theory. The situation is very similar to Yang Mills theory, where the constraint is given by the Gauss law operator which is the generator of time independent gauge transformations. (See e.g. Refs. \cite{Pawlowski98}-\cite{Jo}.)

\end{document}